\documentclass[%
 preprint,
 showpacs,
 showkeys,
 preprintnumbers,
 amsmath,amssymb,
 aip,
 jmp,
  longbibliography,
 ]{revtex4-1}

\usepackage{graphicx}

\newtheorem{theo}{Theorem}
\newtheorem{lemma}{Lemma}

\begin{document}

\title{Density conditions for quantum propositions}

\author{Hans Havlicek}
\affiliation{Institut f\"ur Geometrie, Vienna
    University of Technology, Wiedner Hauptstra\ss e 8-10/1133, A-1040
    Vienna, Austria}
\email{havlicek@geometrie.tuwien.ac.at}
\homepage[]{http://www.dmg.tuwien.ac.at/havlicek/}

\author{Karl Svozil}
\affiliation{Institute of Theoretical Physics, Vienna
    University of Technology, Wiedner Hauptstra\ss e 8-10/136, A-1040
    Vienna, Austria}
\email{svozil@tuwien.ac.at}
\homepage[]{http://tph.tuwien.ac.at/~svozil}


\begin{abstract}
As has already been pointed out by Birkhoff and von Neumann, quantum
logic can be formulated in terms of projective geometry.
In three-dimensional Hilbert space, elementary logical propositions are
associated with one-dimensional subspaces, corresponding to points of
the projective plane.
It is shown that, starting with three such propositions corresponding to
some basis $\{{\vec u},{\vec v},{\vec w}\}$, successive application of
the binary logical operation $(x,y)\mapsto (x\vee y)^\perp$  generates a
set of elementary
propositions which is countable infinite and dense in the projective
plane if and only if
no vector of the basis $\{{\vec u},{\vec v},{\vec w}\}$ is orthogonal to
the other ones.
\end{abstract}

\pacs{02.10.Ab,02.40.-k}
\keywords{quantum logic, generation of projectors, density of quantum observables}

\maketitle

\section{Introduction}
The geometrization of quantum logic was initiated
by Birkhoff and von Neumann \cite{birkhoff-36}.  In their ``top-down''
approach, the logical entities are identified with Hilbert space
entities as follows.
Elementary propositions are identified with one-dimensional subspaces or
with the vector spanning that subspace.  The binary logical operations
``and'' ($\wedge$) and ``or'' ($\vee$) correspond to the set theoretic
intersection and to the linear span, respectively.  The unary logical
operation ``not'' ( $^\perp$ ) corresponds to the orthogonal subspace.
The proposition which is always false is identified with the null
vector.  The proposition which is always true is identified with the
entire Hilbert space.  In that way, the geometry of Hilbert space
induces a logical structure which, if Hilbert space quantum mechanics
\cite{v-neumann-49} is an appropriate theory of quantum physics,
describes correctly the logical structure of measurements (cf. Refs.
\cite{jammer1,pulmannova-91,kalmbach-83,cohen,giuntini-91}).

In what follows, we concentrate on the following question.  Assume we
start with a set $\{u,v,w\}$ of three elementary quantum mechanical
propositions representable as one-dimensional subspaces of
three-dimensional Hilbert space.  New pro\-po\-si\-tions can be formed
from
the old ones by the logical operations ``and, or, not.''  In particular,
the operation $(x\vee y)^\perp$ corresponding to ``not ($x$ or $y$)'' is
just the subspace spanned by the vector product ${\vec x}\times {\vec
y}$.  Suppose this operation is carried out recursively.  That is,
at each step we form the vector product of all (nonparallel) vectors and
add the (nonparallel) results to the previous set of vectors.  One may
ask, what are the conditions for the resulting set (of intersection
points with the unit ball) to be dense?  Evidently, the set of
one-dimensional subspaces spanned by the recursive application of the
vector product can at most be countable (cardinality $\aleph_0$).  It is
less obvious if there can be any regions or ``holes'' formed by the
recursively obtained set of one-dimensional subspaces which are
unreachable. An answer is given in theorem
\ref{theorem-3}.

As has been already pointed out by Birkhoff and von Neumann
\cite{v-neumann-49}, the structure obtained for three-dimensional
Hilbert space is essentially a projective plane.  Points of the
projective geometry are identified with elementary propositions, and
lines are identified with two-dimensional subspaces.
We emphasize this point of view by reformulating the above problem into
the geometric language of the real projective plane endowed with the
elliptic metric.

The original motivation for this question originates from the
consideration of Kochen-Specker type constructions
\cite{kochen1,mermin-93}.  It has been conjectured that every set of
three non\-ortho\-go\-nal one-dimensional subspaces generates a
Kochen-Specker
paradox \cite{svozil-tkadlec}.  More generally, one could ask if any
single elementary proposition (corresponding to a one-dimensional
subspace of three-dimensional Hilbert space) can be approximated by a
logical construction originating from just three propositions
(corresponding to nonorthogonal one-dimensional subspaces of
three-dimensional Hilbert space).

It has to be kept in mind, however, that a consistent two-valued
measure---serving as a classical truth function---will in general not be
definable on the set of recursively generated one-dimensional subspaces
identifiable with elementary propositions.  Indeed, due to
complementarity, even for the generating set of three vectors, such an
identification of truth functions will only have an operational
(physical) meaning if these vectors were mutually orthogonal---a
condition which would yield a trivial orthogonal tripod configuration,
for which any recursion does not produce any additional vectors.

\section{Subplanes of projective planes}\label{PLANES}

A {\em projective plane}\/ is formally a geometric structure $({\cal P},{\cal
L},I)$ consisting of a set ${\cal P}$ of elements called {\em points}, a set
${\cal L}$ of elements called {\em lines}\/ and a binary relation
$I\subset{\cal P}\times{\cal L}$ called {\em incidence}\/ satisfying the
following axioms:

\vspace{1em}

(P1)  Any two distinct points are incident with exactly one common line.

(P2)  Any two distinct lines are incident with a common point.

(P3)  There are four points, no three of which are incident with a common
line.

\vspace{1em}

Instead of $(p,L)\in I$ we also write $p\,I\,L$ and use familiar expressions
like ``$p$ is on $L$'', ``$L$ is running through $p$'' etc. A set of points
is said to be {\em collinear}, if all points are on a common line, a {\em
triangle}\/ is a set of three non-collinear points, a {\em quadrangle}\/ is a
set of four points satisfying the condition of axiom (P3). If we are given
two distinct points $p_1,p_2\in {\cal P}$ then $p_1\vee p_2$ denotes the
unique line joining these two points. By (P1) and (P2), two distinct lines
$L_1,L_2\in {\cal L}$ meet at a unique point which  is written as $L_1\wedge
L_2$. For basic properties of projective planes see \cite[Chapter
4]{Buekenhout-95}, \cite{Hughes-73} or \cite{Stevenson-72}.

Let $F$ be a skewfield (division ring). Then $F^3$ (regarded as left vector
space over $F$) gives rise to a projective plane as follows: Define ${\cal
P}$ as set of all one-dimensional subspaces of $F^3$, viz.
   \begin{equation}
{\cal P}:=\{F{\vec a} \mid {\vec o}\neq {\vec a}\in F^3\},
   \end{equation}
and ${\cal L}$ as the set of all two-dimensional subspaces of $F^3$.
Incidence is defined by
   \begin{equation}
   I:= \{(F{\vec a},L)\in{\cal P}\times{\cal L}
   \mid F{\vec a}\subset L\}.
   \end{equation}
We set $({\cal P},{\cal L}, I)=:\mbox{PG}(2,F)$. See e.g.\ \cite[p.\
29]{Gruenberg-77}, \cite[p.\ 222]{Kostrikin-89} or the textbooks mentioned
above for more details.

We remark that there are also projective planes that are not isomorphic to
any plane of the form $\mbox{PG}(2,F)$. Such projective planes are called
{\em Non-Desarguesian}\/ and will not be of interest in this paper.

Suppose that $({\cal P},{\cal L},I)$ is a projective plane and that
$\widetilde{{\cal P}}$ is any subset of ${\cal P}$. Put
   \begin{equation}\label{Lsub}
   \widetilde{{\cal L}}:=
   \{p_1\vee p_2\mid p_1,p_2\in \widetilde{{\cal P}}, p_1\neq p_2\}
   \mbox{ and }
   \widetilde{I}:=I\cap(\widetilde{{\cal P}}\times \widetilde{{\cal L}}).
\end{equation}
The substructure $(\widetilde{{\cal P}},\widetilde{{\cal L}},\widetilde{I})$
is satisfying axiom (P1), but not necessarily (P2) or (P3). If
$(\widetilde{{\cal P}},\widetilde{{\cal L}},\widetilde{I})$ is a projective
plane, then it is called a {\em projective subplane}\/ of $({\cal P},{\cal
L},I)$. A {\em degenerate subplane}\/ $(\widetilde{{\cal P}},\widetilde{{\cal
L}},\widetilde{I})$ is satisfying (P2), but not (P3).

All degenerate subplanes are easily described: If $\#\widetilde{{\cal L}}\leq
1$, then $\widetilde{{\cal P}}$ is a set of collinear points. If
$\#\widetilde{{\cal L}}\geq 2$, then $\widetilde{{\cal P}}$ is formed by a
set of two or more points on a line, say $L$, plus one more point, say $u$,
off the line $L$. This $L$ is the only line in $\widetilde{{\cal L}}$ not
running through $u$.

In $\mbox{PG}(2,F)$ we may obtain a projective subplane as follows: Let
$\{{\vec b}_1,{\vec b}_2,{\vec b}_3\}\subset F^3$ be a basis and let
$\widetilde{F}\subset F$ be a sub-skewfield of $F$. Then set
   \begin{equation}\label{PSUB}
   \widetilde{{\cal P}}=
   \{ F{\vec a} \mid {\vec a}=\sum_{i=1}^{3}\xi_i{\vec b}_i, (0,0,0)\neq
   (\xi_1,\xi_2,\xi_3)\in \widetilde{F}^3\}
   \end{equation}
and define $\widetilde{{\cal L}},\widetilde{I}$ according to (\ref{Lsub}).
The verification of (P2) amounts to solving a homogeneous system of linear
equations within the sub-skewfield $\widetilde{F}$. A quadrangle in
$\widetilde{{\cal P}}$ is given by
$\{{\Bbb R}{\vec b}_1,{\Bbb R}{\vec b}_2,{\Bbb R}{\vec b}_3,
{\Bbb R}({\vec b}_1+{\vec b}_2+{\vec b}_3)\}$.

The backbone of this article is the following innocently looking result
\cite[p.\ 266]{Stevenson-72}: Any projective subplane of $\mbox{PG}(2,F)$ is
of the form (\ref{PSUB}). See also \cite[p.\ 1008]{Rigby-65}. This allows to
recover an algebraic structure, namely a sub-skewfield of $F$, from a
projective subplane of $\mbox{PG}(2,F)$. Let us add, for the sake of
completeness, the following remark: If in (\ref{PSUB}) the basis $\{{\vec
b}_1,{\vec b}_2,{\vec b}_3\}$ is replaced by $\{\alpha{\vec b}_1,\alpha{\vec
b}_2,\alpha{\vec b}_3\}$ for some non-zero $\alpha\in F$ and if
$\widetilde{F}$ is modified to the sub-skewfield
$\alpha\widetilde{F}\alpha^{-1}$, then $\widetilde{{\cal P}}$ remains
unchanged. Actually, a projective subplane of $\mbox{PG}(2,F)$ determines
``its'' sub-skewfield of $F$ only to within transformation under inner
automorphisms of $F$. Clearly, for a (commutative) field $F$ this means
uniqueness.

We confine our attention to the {\em real projective plane}\/
$\mbox{PG}(2,{\Bbb R})$. The {\em elliptic metric}\/ on ${\cal P}$ is given
by
   \begin{equation}\label{ELL}
   d : {\cal P}\times {\cal P}\rightarrow {\Bbb R}, ({\Bbb R}{\vec a},
   {\Bbb R}{\vec b}) \mapsto
   \arccos\frac{|{\vec a}\cdot{\vec b}|}
               {\|{\vec a}\|\,\|{\vec b}\|}
   \in \left[0,\frac{\pi}{2}\right],
   \end{equation}
where $\cdot$ denotes the standard dot product and $\|\;\|$ stands for the
Euclidean norm of ${\Bbb R}^3$. The {\em elliptic distance}\/ $d({\Bbb
R}{\vec a},{\Bbb R}{\vec b})$ of two points of $\mbox{PG}(2,{\Bbb R})$ is
just the Euclidean angle of the corresponding one-dimensional subspaces
through the origin of ${\Bbb R}^3$.
It  is invariant under transformations
(e.g., rotations) which preserve normality. Besides, a connection can be
made between the elliptic distance and the more physically motivated
{\em statistical distance} \cite{wootters-81}.

For each point ${\Bbb R}{\vec a}$ of $\mbox{PG}(2,{\Bbb R})$ there are
exactly two unit vectors in ${\Bbb R}{\vec a}$. This gives the well-known
alternative description of the real projective plane: The ``points'' may be
viewed as unordered pairs of opposite points of the unit sphere, the
``lines'' are the great circles and incidence is defined via inclusion. In
this interpretation the elliptic distance is equal to the {\em spherical
distance}\/ \cite[Chapter VI]{Coxeter-78}.

If $T$ is a subset of ${\Bbb R}^3$ then $T^\perp := \{{\vec a}\mid{\vec
a}\cdot{\vec t}=0\mbox{ for all }{\vec t}\in T\}$ is a subspace. In geometric
terms $\perp$ is a {\em polarity}\/ of the projective plane
$\mbox{PG}(2,{\Bbb R})$; cf.\ \cite[Chapter 17]{Buekenhout-95}, \cite[p.\
52]{Coxeter-78}, \cite[p.\ 110]{Gruenberg-77} or \cite[p.\ 45]{Hughes-73}.
Points and lines are interchanged bijectively subject to the rule ${\Bbb
R}{\vec a}\,(\in{\cal P})\mapsto{\vec a}^\perp\,(\in{\cal L})$. The geometric
operations of ``join'' ($\vee$) and ``meet'' ($\wedge$) therefore allow a
simple algebraic description: Given linearly independent vectors ${\vec a},
{\vec b}\in{\Bbb R}^3$ then
   \begin{equation}\label{VERBINDEN}
   {\Bbb R}{\vec a}\vee{\Bbb R}{\vec b} =
   ({\vec a}\times{\vec b})^\perp,
   \end{equation}
   \begin{equation}\label{SCHNEIDEN}
   {\vec a}^\perp\wedge{\vec b}^\perp =
   {\Bbb R}({\vec a}\times{\vec b}).
   \end{equation}

The following result is essentially $(\widetilde{F}={\Bbb Q})$ due to A.F.
M\"obius:

   \begin{lemma}
   If $(\widetilde{{\cal P}},\widetilde{{\cal L}},\widetilde{I})$ is a
   projective subplane of $({\cal P},{\cal L},I) = \mbox{{\rm PG}}(2,{\Bbb
   R})$, then $\widetilde{{\cal P}}$ is dense in ${\cal P}$.
   \end{lemma}
{\em Proof.} Let $\widetilde{{\cal P}}$ be given according to (\ref{PSUB})
with $\widetilde{F}\subset {\Bbb R}$. The field ${\Bbb Q}$ of rational
numbers equals the intersection of all subfields of ${\Bbb R}$, whence ${\Bbb
Q}\subset \widetilde{F}$. Given a point ${\Bbb R}{\vec a}\in{\cal P}$ we
obtain
   \begin{equation}
   {\vec a} = \xi_1{\vec b}_1 + \xi_2{\vec b}_2 + \xi_3{\vec b}_3
   \mbox{ with }
   (\xi_1,\xi_2,\xi_3)\in {\Bbb R}^3.
   \end{equation}
There exist three sequences
   \begin{equation}
   (\xi_{j,i})_{i\in{{\Bbb N}}},
   \mbox{ with }
   \xi_{j,i}\in {\Bbb Q}\setminus\{0\}
   \mbox { and }
   \lim_{i\to\infty}\xi_{j,i}= \xi_j\; (j\in\{1,2,3\}).
   \end{equation}
Defining
   \begin{equation}
   {\vec a}_i :=
   \xi_{1,i}{\vec b}_1 + \xi_{2,i}{\vec b}_2 +\xi_{3,i}{\vec b}_3
   \neq {\vec o}\; (i\in{\Bbb N})
   \end{equation}
yields a sequence of points ${\Bbb R}{\vec a}_i\in\widetilde{{\cal P}}$ with
$({\Bbb R}{\vec a}_i)_{i\in{{\Bbb N}}} \to {\Bbb R}{\vec a}$, since, by the
continuity of dot product and norm,
   \begin{equation}
   \lim_{i\to\infty}
   \frac{{\vec a}\cdot{\vec a}_i}{\|{\vec a}\|\,\|{\vec a}_i\|}
   =
   \frac{{\vec a}\cdot{\vec a}}{\|{\vec a}\|\,\|{\vec a}\|}
   = 1.
   \end{equation}
This completes the proof. $\Box$

\vspace{1em}

The projective subplanes of $\mbox{PG}(2,{\Bbb R})$ belonging to the rational
number field are called {\em M\"obius nets}\/. They allow a simple recursive
geometric construction \cite[p.\ 140]{Klein-68}: Starting with a quadrangle
one draws all the lines spanned by these points. Next mark all points of
intersection arising from these lines. With this set of points the procedure
is repeated, and so on. The set of all points that can be reached in a finite
number of steps gives then a projective subplane over ${\Bbb Q}$.

\section{Main theorems}

   \begin{theo}\label{TH1}
   Let $V_1=\{{\vec u},{\vec v},{\vec w}\}$ be a basis of ${\Bbb R}^3$.
   Define subsets $V_i, V$ of ${\Bbb R}^3$ as follows:
      \begin{equation}
      V_{i+1}:=
      V_i\cup
      \{{\vec r}\times{\vec s}\mid{\vec r},{\vec s}\in V_i,\;
      {\vec r}\times{\vec s}\neq{\vec o}\}\:
      (i\in{\Bbb N}), \quad
      V:=\bigcup_{i=1}^\infty V_i.
      \end{equation}
   Then
      \begin{equation}\label{PSUB-V}
      \widetilde{{\cal P}}:=\{{\Bbb R}{\vec a}\mid{\vec a}\in V\}
      \end{equation}
   yields a projective or degenerate subplane
   $(\widetilde{{\cal P}},\widetilde{{\cal L}},\widetilde{I})$
   of $\mbox{{\rm PG}}(2,{\Bbb R})$
which is ortho-closed.
That is,
   ${\Bbb R}{\vec a}\in\widetilde{{\cal P}}$ implies
   ${\vec a}^\perp\in\widetilde{{\cal L}}$.
  \end{theo}

{\em Proof.} Let $L_1,L_2\in\widetilde{{\cal L}}$ be distinct. By
(\ref{VERBINDEN}) and the definition of $\widetilde{{\cal L}}$, there are
vectors ${\vec p}_1,{\vec q}_1,{\vec p}_2,{\vec q}_2\in V$ with
\begin{equation}
   L_1=({\vec p}_1\times{\vec q}_1)^\perp,\quad
   L_2=({\vec p}_2\times{\vec q}_2)^\perp.
   \end{equation}
Now (\ref{SCHNEIDEN}) yields
   \begin{equation}
   L_1\wedge L_2 =
   {\Bbb R}(({\vec p}_1\times{\vec q}_1)\times({\vec p}_2\times{\vec q}_2))
   \in\widetilde{{\cal P}}.
   \end{equation}
This establishes (P2).

Given a point ${\Bbb R}{\vec a}\in\widetilde{{\cal P}}$, there exist two
vectors in $V_1$, say ${\vec u},{\vec v}$, such that $\{{\vec a},{\vec u},
{\vec v}\}$ is a basis of ${\Bbb R}^3$. Then $u\notin\mbox{span}\,\{{\vec a},
{\vec v}\}=({\vec a}\times{\vec v})^\perp$, but ${\vec u}\in({\vec a} \times
{\vec u})^\perp$. Thus ${\Bbb R}({\vec a}\times{\vec v})$ and ${\Bbb R}({\vec
a} \times{\vec v})$ are distinct points of $\widetilde{{\cal P}}$ on the line
${\vec a}^\perp$. $\Box$

Observe that axiom (P2) may be derived alternatively from the well-known
formula
   \begin{equation}
      \begin{array}{rcl}
      ({\vec p}_1\times{\vec q}_1)\times({\vec p}_2\times{\vec q}_2)
      & = &
       \det({\vec p}_1,{\vec q}_1,{\vec q}_2) {\vec p}_2
      -\det({\vec p}_1,{\vec q}_1,{\vec p}_2) {\vec q}_2
      \\
      {}
      & = &
       \det({\vec p}_1,{\vec p}_2,{\vec q}_2) {\vec q}_1
      -\det({\vec q}_1,{\vec p}_2,{\vec q}_2) {\vec p}_1,

   \end{array}
   \end{equation}
since linearly dependent vectors yield collinear points.

   \begin{theo}
   The subplane $(\widetilde{{\cal P}},\widetilde{{\cal L}},\widetilde{I})$
   described in Theorem \ref{TH1} is degenerate if and only if one vector of
   the basis $\{{\vec u},{\vec v},{\vec w}\}$ is orthogonal to the other
   ones.
   \end{theo}
{\em Proof.} Let $(\widetilde{{\cal P}},\widetilde{{\cal L}},\widetilde{I})$
be degenerate. $\{{\Bbb R}{\vec u},{\Bbb R}{\vec v},{\Bbb R}{\vec w}\}$ being
a triangle forces $\#\widetilde{{\cal L}}\geq 3$. We read off from the
description of degenerate subplanes in section \ref{PLANES} that
$\widetilde{{\cal P}}$ has to consist of one point of this triangle, say
${\Bbb R}{\vec u}$, and a subset of points on the line joining ${\Bbb R}{\vec
v}$ and ${\Bbb R}{\vec w}$. The line ${\vec u}^\perp$ belongs to
$\widetilde{{\cal L}}$ by Theorem \ref{TH1}. Now ${\vec u}\notin {\vec
u}^\perp$ tells us that the point ${\Bbb R}{\vec u}$ is off that line. Since
${\Bbb R}{\vec u}$ is on all lines of $\widetilde{{\cal L}}$ but one, we
obtain ${\vec v},{\vec w}\in{\vec u}^\perp$.

Conversely, assume that ${\vec v},{\vec w}\in{\vec u}^\perp$. Then
   \begin{equation}
   \widetilde{{\cal P}} =
   \{{\Bbb R}{\vec u},{\Bbb R}{\vec v},{\Bbb R}{\vec w},{\Bbb R}({\vec
   u}\times{\vec v}), {\Bbb R}({\vec u}\times{\vec w})\}
   \end{equation}
is a set of five points if ${\vec v}\not\perp{\vec w}$, and it is a set of
just three points if ${\vec u},{\vec v},{\vec w}$ are mutually orthogonal.
Thus $\widetilde{{\cal P}}$ yields a degenerate subplane. $\Box$

\vspace{1em}

Summing up, gives this final result:

   \begin{theo}
\label{theorem-3}
   With the settings of Theorem \ref{TH1} the following assertions are
   equivalent:

   1. The basis $\{{\vec u},{\vec v},{\vec w}\}$ of ${\Bbb R}^3$ does not
   contain a vector that is orthogonal to the remaining ones.

   2. The point set $\widetilde{\cal P}$ given by (\ref{PSUB-V}) is dense in
   $\mbox{\rm PG}(2,\Bbb R)$.

   3. The point set $\widetilde{\cal P}$ given by (\ref{PSUB-V}) is infinite.
   \end{theo}


\begin{thebibliography}{21}%
\makeatletter
\providecommand \@ifxundefined [1]{%
 \@ifx{#1\undefined}
}%
\providecommand \@ifnum [1]{%
 \ifnum #1\expandafter \@firstoftwo
 \else \expandafter \@secondoftwo
 \fi
}%
\providecommand \@ifx [1]{%
 \ifx #1\expandafter \@firstoftwo
 \else \expandafter \@secondoftwo
 \fi
}%
\providecommand \natexlab [1]{#1}%
\providecommand \enquote  [1]{``#1''}%
\providecommand \bibnamefont  [1]{#1}%
\providecommand \bibfnamefont [1]{#1}%
\providecommand \citenamefont [1]{#1}%
\providecommand \href@noop [0]{\@secondoftwo}%
\providecommand \href [0]{\begingroup \@sanitize@url \@href}%
\providecommand \@href[1]{\@@startlink{#1}\@@href}%
\providecommand \@@href[1]{\endgroup#1\@@endlink}%
\providecommand \@sanitize@url [0]{\catcode `\\12\catcode `\$12\catcode
  `\&12\catcode `\#12\catcode `\^12\catcode `\_12\catcode `\%12\relax}%
\providecommand \@@startlink[1]{}%
\providecommand \@@endlink[0]{}%
\providecommand \url  [0]{\begingroup\@sanitize@url \@url }%
\providecommand \@url [1]{\endgroup\@href {#1}{\urlprefix }}%
\providecommand \urlprefix  [0]{URL }%
\providecommand \Eprint [0]{\href }%
\providecommand \doibase [0]{http://dx.doi.org/}%
\providecommand \selectlanguage [0]{\@gobble}%
\providecommand \bibinfo  [0]{\@secondoftwo}%
\providecommand \bibfield  [0]{\@secondoftwo}%
\providecommand \translation [1]{[#1]}%
\providecommand \BibitemOpen [0]{}%
\providecommand \bibitemStop [0]{}%
\providecommand \bibitemNoStop [0]{.\EOS\space}%
\providecommand \EOS [0]{\spacefactor3000\relax}%
\providecommand \BibitemShut  [1]{\csname bibitem#1\endcsname}%
\let\auto@bib@innerbib\@empty
\bibitem [{\citenamefont {Birkhoff}\ and\ \citenamefont {{von
  Neumann}}(1936)}]{birkhoff-36}%
  \BibitemOpen
  \bibfield  {author} {\bibinfo {author} {\bibfnamefont {G.}~\bibnamefont
  {Birkhoff}}\ and\ \bibinfo {author} {\bibfnamefont {J.}~\bibnamefont {{von
  Neumann}}},\ }\bibfield  {title} {\enquote {\bibinfo {title} {The logic of
  quantum mechanics},}\ }\href {\doibase 10.2307/1968621} {\bibfield  {journal}
  {\bibinfo  {journal} {Annals of Mathematics}\ }\textbf {\bibinfo {volume}
  {37}},\ \bibinfo {pages} {823--843} (\bibinfo {year} {1936})}\BibitemShut
  {NoStop}%
\bibitem [{\citenamefont {{von Neumann}}(1932)}]{v-neumann-49}%
  \BibitemOpen
  \bibfield  {author} {\bibinfo {author} {\bibfnamefont {J.}~\bibnamefont {{von
  Neumann}}},\ }\href@noop {} {\emph {\bibinfo {title} {{M}athematische
  {G}rundlagen der {Q}uantenmechanik}}}\ (\bibinfo  {publisher} {Springer},\
  \bibinfo {address} {Berlin},\ \bibinfo {year} {1932})\ \bibinfo {note}
  {{E}nglish translation in Ref.~\cite{v-neumann-55}}\BibitemShut {NoStop}%
\bibitem [{\citenamefont {Jammer}(1974)}]{jammer1}%
  \BibitemOpen
  \bibfield  {author} {\bibinfo {author} {\bibfnamefont {M.}~\bibnamefont
  {Jammer}},\ }\href@noop {} {\emph {\bibinfo {title} {The Philosophy of
  Quantum Mechanics}}}\ (\bibinfo  {publisher} {John Wiley \& Sons},\ \bibinfo
  {address} {New York},\ \bibinfo {year} {1974})\BibitemShut {NoStop}%
\bibitem [{\citenamefont {Pt{\'{a}}k}\ and\ \citenamefont
  {Pulmannov{\'{a}}}(1991)}]{pulmannova-91}%
  \BibitemOpen
  \bibfield  {author} {\bibinfo {author} {\bibfnamefont {P.}~\bibnamefont
  {Pt{\'{a}}k}}\ and\ \bibinfo {author} {\bibfnamefont {S.}~\bibnamefont
  {Pulmannov{\'{a}}}},\ }\href@noop {} {\emph {\bibinfo {title} {Orthomodular
  Structures as Quantum Logics}}}\ (\bibinfo  {publisher} {Kluwer Academic
  Publishers},\ \bibinfo {address} {Dordrecht},\ \bibinfo {year}
  {1991})\BibitemShut {NoStop}%
\bibitem [{\citenamefont {Kalmbach}(1983)}]{kalmbach-83}%
  \BibitemOpen
  \bibfield  {author} {\bibinfo {author} {\bibfnamefont {G.}~\bibnamefont
  {Kalmbach}},\ }\href@noop {} {\emph {\bibinfo {title} {Orthomodular
  Lattices}}}\ (\bibinfo  {publisher} {Academic Press},\ \bibinfo {address}
  {New York},\ \bibinfo {year} {1983})\BibitemShut {NoStop}%
\bibitem [{\citenamefont {Cohen}(1989)}]{cohen}%
  \BibitemOpen
  \bibfield  {author} {\bibinfo {author} {\bibfnamefont {D.~W.}\ \bibnamefont
  {Cohen}},\ }\href@noop {} {\emph {\bibinfo {title} {An Introduction to
  Hilbert Space and Quantum Logic}}}\ (\bibinfo  {publisher} {Springer},\
  \bibinfo {address} {New York},\ \bibinfo {year} {1989})\BibitemShut {NoStop}%
\bibitem [{\citenamefont {Giuntini}(1991)}]{giuntini-91}%
  \BibitemOpen
  \bibfield  {author} {\bibinfo {author} {\bibfnamefont {R.}~\bibnamefont
  {Giuntini}},\ }\href@noop {} {\emph {\bibinfo {title} {Quantum Logic and
  Hidden Variables}}}\ (\bibinfo  {publisher} {BI Wissenschaftsverlag},\
  \bibinfo {address} {Mannheim},\ \bibinfo {year} {1991})\BibitemShut {NoStop}%
\bibitem [{\citenamefont {Kochen}\ and\ \citenamefont
  {Specker}(1967)}]{kochen1}%
  \BibitemOpen
  \bibfield  {author} {\bibinfo {author} {\bibfnamefont {S.}~\bibnamefont
  {Kochen}}\ and\ \bibinfo {author} {\bibfnamefont {E.~P.}\ \bibnamefont
  {Specker}},\ }\bibfield  {title} {\enquote {\bibinfo {title} {The problem of
  hidden variables in quantum mechanics},}\ }\href {\doibase
  10.1512/iumj.1968.17.17004} {\bibfield  {journal} {\bibinfo  {journal}
  {Journal of Mathematics and Mechanics (now Indiana University Mathematics
  Journal)}\ }\textbf {\bibinfo {volume} {17}},\ \bibinfo {pages} {59--87}
  (\bibinfo {year} {1967})},\ \bibinfo {note} {reprinted in Ref.~\cite[pp.
  235--263]{specker-ges}}\BibitemShut {NoStop}%
\bibitem [{\citenamefont {Mermin}(1993)}]{mermin-93}%
  \BibitemOpen
  \bibfield  {author} {\bibinfo {author} {\bibfnamefont {N.~D.}\ \bibnamefont
  {Mermin}},\ }\bibfield  {title} {\enquote {\bibinfo {title} {Hidden variables
  and the two theorems of {J}ohn {B}ell},}\ }\href {\doibase
  10.1103/RevModPhys.65.803} {\bibfield  {journal} {\bibinfo  {journal}
  {Reviews of Modern Physics}\ }\textbf {\bibinfo {volume} {65}},\ \bibinfo
  {pages} {803--815} (\bibinfo {year} {1993})}\BibitemShut {NoStop}%
\bibitem [{\citenamefont {Svozil}\ and\ \citenamefont
  {Tkadlec}(1996)}]{svozil-tkadlec}%
  \BibitemOpen
  \bibfield  {author} {\bibinfo {author} {\bibfnamefont {K.}~\bibnamefont
  {Svozil}}\ and\ \bibinfo {author} {\bibfnamefont {J.}~\bibnamefont
  {Tkadlec}},\ }\bibfield  {title} {\enquote {\bibinfo {title} {Greechie
  diagrams, nonexistence of measures in quantum logics and {K}ochen--{S}pecker
  type constructions},}\ }\href {\doibase 10.1063/1.531710} {\bibfield
  {journal} {\bibinfo  {journal} {Journal of Mathematical Physics}\ }\textbf
  {\bibinfo {volume} {37}},\ \bibinfo {pages} {5380--5401} (\bibinfo {year}
  {1996})}\BibitemShut {NoStop}%
\bibitem [{\citenamefont {Buekenhout}(1995)}]{Buekenhout-95}%
  \BibitemOpen
  \bibinfo {editor} {\bibfnamefont {F.}~\bibnamefont {Buekenhout}},\ ed.,\
  \href@noop {} {\emph {\bibinfo {title} {Handbook of Incidence Geometry}}}\
  (\bibinfo  {publisher} {Elsevier},\ \bibinfo {address} {Amsterdam},\ \bibinfo
  {year} {1995})\BibitemShut {NoStop}%
\bibitem [{\citenamefont {Hughes}\ and\ \citenamefont
  {Piper}(1973)}]{Hughes-73}%
  \BibitemOpen
  \bibfield  {author} {\bibinfo {author} {\bibfnamefont {D.~R.}\ \bibnamefont
  {Hughes}}\ and\ \bibinfo {author} {\bibfnamefont {F.~C.}\ \bibnamefont
  {Piper}},\ }\href@noop {} {\emph {\bibinfo {title} {Projective Planes}}}\
  (\bibinfo  {publisher} {Springer},\ \bibinfo {address} {New
  York,Heidelberg,Berlin},\ \bibinfo {year} {1973})\BibitemShut {NoStop}%
\bibitem [{\citenamefont {Stevenson}(1972)}]{Stevenson-72}%
  \BibitemOpen
  \bibfield  {author} {\bibinfo {author} {\bibfnamefont {F.~W.}\ \bibnamefont
  {Stevenson}},\ }\href@noop {} {\emph {\bibinfo {title} {Projective Planes}}}\
  (\bibinfo  {publisher} {Freeman},\ \bibinfo {address} {San Francisco},\
  \bibinfo {year} {1972})\BibitemShut {NoStop}%
\bibitem [{\citenamefont {Gruenberg}\ and\ \citenamefont
  {Weir}(1977)}]{Gruenberg-77}%
  \BibitemOpen
  \bibfield  {author} {\bibinfo {author} {\bibfnamefont {K.~W.}\ \bibnamefont
  {Gruenberg}}\ and\ \bibinfo {author} {\bibfnamefont {A.~J.}\ \bibnamefont
  {Weir}},\ }\href@noop {} {\emph {\bibinfo {title} {Linear Geometry (2nd
  Edition)}}}\ (\bibinfo  {publisher} {Springer},\ \bibinfo {address} {New
  York, Heidelberg, Berlin},\ \bibinfo {year} {1977})\BibitemShut {NoStop}%
\bibitem [{\citenamefont {Kostrikin}\ and\ \citenamefont
  {Manin}(1989)}]{Kostrikin-89}%
  \BibitemOpen
  \bibfield  {author} {\bibinfo {author} {\bibfnamefont {A.~I.}\ \bibnamefont
  {Kostrikin}}\ and\ \bibinfo {author} {\bibfnamefont {Y.~I.}\ \bibnamefont
  {Manin}},\ }\href@noop {} {\emph {\bibinfo {title} {Linear Algebra and
  Geometry}}}\ (\bibinfo  {publisher} {Gordon and Breach},\ \bibinfo {address}
  {New York -- London -- Paris},\ \bibinfo {year} {1989})\BibitemShut {NoStop}%
\bibitem [{\citenamefont {Rigby}(1965)}]{Rigby-65}%
  \BibitemOpen
  \bibfield  {author} {\bibinfo {author} {\bibfnamefont {J.~F.}\ \bibnamefont
  {Rigby}},\ }\bibfield  {title} {\enquote {\bibinfo {title} {Affine subplanes
  of finite projective planes},}\ }\href@noop {} {\bibfield  {journal}
  {\bibinfo  {journal} {Can. Journal Math.}\ }\textbf {\bibinfo {volume}
  {17}},\ \bibinfo {pages} {977--1014} (\bibinfo {year} {1965})}\BibitemShut
  {NoStop}%
\bibitem [{\citenamefont {Wootters}(1981)}]{wootters-81}%
  \BibitemOpen
  \bibfield  {author} {\bibinfo {author} {\bibfnamefont {W.~K.}\ \bibnamefont
  {Wootters}},\ }\bibfield  {title} {\enquote {\bibinfo {title} {Statistical
  distance and {H}ilbert space},}\ }\href@noop {} {\bibfield  {journal}
  {\bibinfo  {journal} {Physical Review D}\ }\textbf {\bibinfo {volume} {23}},\
  \bibinfo {pages} {357--362} (\bibinfo {year} {1981})}\BibitemShut {NoStop}%
\bibitem [{\citenamefont {Coxeter}(1978)}]{Coxeter-78}%
  \BibitemOpen
  \bibfield  {author} {\bibinfo {author} {\bibfnamefont {H.~S.~M.}\
  \bibnamefont {Coxeter}},\ }\href@noop {} {\emph {\bibinfo {title}
  {Non-Euclidean Geometry}}}\ (\bibinfo  {publisher} {University of Toronto
  Press},\ \bibinfo {address} {Toronto -- Buffalo -- London},\ \bibinfo {year}
  {1978})\BibitemShut {NoStop}%
\bibitem [{\citenamefont {Klein}(1968)}]{Klein-68}%
  \BibitemOpen
  \bibfield  {author} {\bibinfo {author} {\bibfnamefont {F.}~\bibnamefont
  {Klein}},\ }\href@noop {} {\emph {\bibinfo {title} {Vorlesungen {\"{u}}ber
  h{\"{o}}here Geometrie, Grundlehren Bd. 22}}}\ (\bibinfo  {publisher}
  {Springer},\ \bibinfo {address} {Berlin -- Heidelberg},\ \bibinfo {year}
  {1968})\BibitemShut {NoStop}%
\bibitem [{\citenamefont {{von Neumann}}(1955)}]{v-neumann-55}%
  \BibitemOpen
  \bibfield  {author} {\bibinfo {author} {\bibfnamefont {J.}~\bibnamefont {{von
  Neumann}}},\ }\href@noop {} {\emph {\bibinfo {title} {Mathematical
  Foundations of Quantum Mechanics}}}\ (\bibinfo  {publisher} {Princeton
  University Press},\ \bibinfo {address} {Princeton, NJ},\ \bibinfo {year}
  {1955})\BibitemShut {NoStop}%
\bibitem [{\citenamefont {Specker}(1990)}]{specker-ges}%
  \BibitemOpen
  \bibfield  {author} {\bibinfo {author} {\bibfnamefont {E.}~\bibnamefont
  {Specker}},\ }\href@noop {} {\emph {\bibinfo {title} {Selecta}}}\ (\bibinfo
  {publisher} {Birkh{\"{a}}user Verlag},\ \bibinfo {address} {Basel},\ \bibinfo
  {year} {1990})\BibitemShut {NoStop}%
\end{thebibliography}

%

\end{document}